\documentclass{article}

\PassOptionsToPackage{numbers,sort&compress}{natbib}
\usepackage[preprint]{neurips_2026}
\usepackage[utf8]{inputenc}
\usepackage[T1]{fontenc}
\usepackage{amsmath,amssymb,amsfonts,amsthm,mathtools}
\usepackage{booktabs}
\usepackage{graphicx}
\usepackage{tabularx}
\usepackage{enumitem}
\usepackage{tikz}
\usetikzlibrary{arrows.meta,positioning,fit,backgrounds}
\usepackage{hyperref}
\usepackage{microtype}
\usepackage{xcolor}
\usepackage{multirow}
\setlist[itemize]{topsep=2pt,itemsep=1pt,parsep=0pt,partopsep=0pt}
\setlist[enumerate]{topsep=2pt,itemsep=1pt,parsep=0pt,partopsep=0pt}

\long\def\comment#1{}
\def\ie{$i.e.$}

\newcolumntype{L}[1]{>{\raggedright\arraybackslash}p{#1}}
\newcolumntype{Y}{>{\raggedright\arraybackslash}X}

\newcommand{\SourceA}{Source~A}
\newcommand{\SourceB}{Source~B}
\newcommand{\SourceC}{Source~C}

\definecolor{colA}{RGB}{220,80,60}
\definecolor{colB}{RGB}{40,100,180}
\definecolor{colC}{RGB}{50,150,80}
\definecolor{colABC}{RGB}{140,60,160}

\usepackage{wrapfig}
\usepackage{sidecap}
\sidecaptionvpos{figure}{t}
\usepackage{lipsum} 
\usepackage{caption}

\title{The Authorization-Execution Gap Is a Major\\
Safety and Security Problem in Open-World Agents}

\author{
\textbf{Baoyuan Wu}\textsuperscript{\rm 1}\thanks{Corresponding Author} \quad
\textbf{Qingshan Liu}\textsuperscript{\rm 2} \quad
\textbf{Adel Bibi}\textsuperscript{\rm 3} \quad
\textbf{Irwin King}\textsuperscript{\rm 4} \quad
\textbf{Siwei Lyu}\textsuperscript{\rm 5} \quad
 \\
\textsuperscript{1}Chinese University of Hong Kong, Shenzhen, China \\
\textsuperscript{2}Nanjing University of Posts and Telecommunications, China \quad \\
\textsuperscript{3}University of Oxford, UK \quad \\
\textsuperscript{4}Chinese University of Hong Kong, China \quad \\
\textsuperscript{5}State University of New York at Buffalo, USA \\
\texttt{wubaoyuan@cuhk.edu.cn; qsliu@nuist.edu.cn; adel.bibi@eng.ox.ac.uk;} \\
\texttt{king@cse.cuhk.edu.hk; siweilyu@buffalo.edu }
}

\begin{document}
\maketitle

\begin{abstract}
This position paper argues that the
\textbf{Authorization-Execution Gap (AEG)} is a
major safety and security problem in open-world
agents. The AEG is the divergence between what a principal
intends to authorize and what an open-world agent
ultimately executes. Because such agents act autonomously across tools, persistent state, and multi-agent
handoffs, even small instances of authorization divergence can cause harm that is difficult or impossible to undo.
We argue that many observed agent failures can be traced to three structural sources of AEG: delegation-level
incompleteness, channel-level corruption, and
composition-level fragmentation. The same observed failure may arise from any of these sources. Without identifying the source, a defense targeting the symptom alone cannot
address the underlying cause. Agent safety and
security should therefore emphasize
source-oriented diagnosis and defense.
Because the structural sources of AEG arise
dynamically during execution, this approach
necessarily requires authorization integrity
checks applied during execution, rather than relying solely on one-shot upfront filtering or post-hoc audit.
For NeurIPS, the implication is that papers on
open-world agents should report not only
outcome-level metrics such as task success or
attack resistance, but also process-level
evidence showing where AEG was detected, constrained, and attributed to a structural source during execution.
\end{abstract}

\section{Introduction}
\label{sec:intro}

\subsection{The Fundamental Shift: From Output Generation to Delegated Execution}
\label{sec:shift}

In large language models (LLMs), the human remains the executor: the model advises, the human decides and acts. In agentic systems, principals delegate bounded mandates to agents, which then decide and act autonomously across sequences of steps that may be costly or impossible to reverse. This shift is fundamental, not a matter of degree. This paper is therefore not about generic LLM safety in text-only settings, where the model produces outputs for a human to evaluate. It is about the distinct safety and security problem that arises when the model becomes an actor: it can use tools, write persistent state, and delegate work across time or across other agents.

\textbf{Our position is that the Authorization-Execution Gap (AEG) is a major safety and security problem in open-world agents. The AEG is the divergence between what a principal intends to authorize and what an open-world agent ultimately executes. Many observed agent failures can be traced to three structural sources of AEG: delegation-level incompleteness, channel-level corruption, and
composition-level fragmentation. Agent safety and security should therefore emphasize source-oriented diagnosis and defense, together with authorization integrity checks during execution.}

Principals do not authorize isolated outputs; they
authorize bounded actions in a workflow. That
authorization is conveyed through delegated context
and then carried into execution, where tool outputs,
state updates, and downstream handoffs may themselves
become decision-guiding. The central question is
whether realized execution stays within the
authorization scope established by delegation.
AEG appears along this path when execution diverges
from what the principal intended to authorize under
pressure from execution-time inputs, persistent
state, and multi-agent composition.

\subsection{Structural Sources of the Gap and Why It Is Dangerous}
\label{sec:irreducible}

AEG is hard to avoid in open-world agents because
it arises from three structural sources.
\textbf{First,} the principal's intended
authorization must be conveyed through delegated
context, which serves only as an incomplete
operational proxy.
As context is often expressed partly in
natural language, boundary cases, exceptions,
escalation conditions, and stopping rules may
remain implicit or ambiguous.
\textbf{Second,} execution channels are corruptible:
tool outputs, web content, and stored state can
redirect later steps when environmental content is
treated as if it carried delegated authority.
\textbf{Third,}
authorization can fragment across stages, tools,
and agents, because locally acceptable steps do not
by themselves guarantee end-to-end authorization
coherence.

\paragraph{Why autonomous execution makes the gap
dangerous.}
In single-turn LLMs, an authorization gap usually
produces incorrect text: the human can reject it,
reinterpret it, or simply do nothing. In agentic
systems, the same gap produces wrong \emph{actions},
persistent state changes, or downstream handoffs
that may have already taken effect before a human
notices. Two execution
properties amplify the gap into harm:
\begin{itemize}[leftmargin=1.2em]
  \item \textbf{Irreversibility}: undoing costs more
    than doing; some actions cannot be undone.
  \item \textbf{Unpredictability}: full consequences
    cannot be foreseen, compounded by emergent
    multi-agent effects.
\end{itemize}
This is why agent safety is not just LLM safety,
but more so. Once a model can act within the
authorization scope established by delegation, the
core issue is no longer only whether an output is
wrong, but whether an autonomous system executes
beyond what the principal intended to authorize.

\subsection{Why This Matters Now}
\label{sec:runtime_necessary}

\textbf{Empirically, the problem is already visible.} Recent agent benchmarks and attack studies document indirect prompt injection, memory poisoning, and multi-agent coordination failures in realistic tool-using systems~\citep{greshake2023notsigned, zhan2024injecagent,debenedetti2024agentdojo, zhang2024asb,chen2024agentpoison,mao2024ibgp, tian2025rethinkmultiagent,ruan2024sandbox, shapira2026agentsofchaos}. The point is not merely that open-world delegation creates new attack names. It is that these failures already arise in realistic agent workflows, and that attack-specific or surface-level descriptions do not by themselves explain where the underlying authorization problem entered. Therefore, a useful response must distinguish whether the failure came from ambiguity in what was delegated, corruption through execution channels, or loss of constraints across multi-agent composition, because those cases call for different diagnoses and different defenses.

\textbf{Existing lines of work are relevant but not sufficient.} The paper's claim is that AEG is the right analytical object for understanding a major class of agent safety and security failures, including prompt injection, persistent corruption, and recomposition failure as different structural manifestations of the same underlying problem rather than unrelated attack families. Existing attack and benchmark papers matter here mainly as stakes evidence and observed-failure evidence: they show that the problem is real, but they do not by themselves explain where the underlying authorization problem entered. 
Threat taxonomies and practitioner guidance \citep{mitre_atlas,owasp_ai_agent_security}, and broader frameworks for agent safety and security \citep{allegrini2025formalizing,ghosh2025safety,arora2025multilayer} help organize the literature, but mainly at the level of attacks, controls, or broad system risks rather than at the level of authorization divergence and its structural sources. Classical security abstractions such as reference monitoring, confinement, least privilege, and information-flow control remain relevant~\citep{saltzer1975protection,lampson1973confinement,sabelfeld2003language}, but open-world delegated workflows place them in a harder setting: authorization is partly expressed in natural language, can be reshaped by newly encountered content, and can weaken across memory and multi-agent composition. Alignment methods such as RLHF~\citep{ouyang2022instructgpt} and Constitutional AI~\citep{bai2022constitutionalai} are also relevant, but they mainly seek to improve model alignment as a route to safer behavior. They do not by themselves remove the need to preserve authorization integrity during execution.

\textbf{For NeurIPS, the paper's contribution is more specific.}
The argument developed in this paper has three
parts: it clarifies how authorization moves from
delegation to execution, shows that the same
structural sources can be traced across documented
agent failures, and derives a source-oriented
account of what should be diagnosed, defended
against, checked during execution, and reported.
If this position is right, agent benchmarks and
system papers should not treat task success or
isolated attack results as sufficient safety
evidence. They should also report how
AEG is detected, constrained, and attributed to
its structural source during execution.

\textbf{Our assumptions and research scope.} 
This paper assumes a benign principal and a trusted
authorization infrastructure. Within that setting,
it studies how realized execution can diverge from
the principal's intended task and intended
authorization scope in a single delegated workflow
or bounded multi-agent system. The analysis excludes
three categories of cases: \textit{harmful authorization}, as
in explicitly harmful-task settings, such as
\textsc{SafeArena}~\citep{tur2025safearena}, \textsc{AgentHarm}~\citep{andriushchenko2025agentharm}, and \textsc{PentestGPT}~\citep{deng2023pentestgpt};
\textit{compromised authorization infrastructure}, such as
tampering with the policy store, provenance
substrate, or checking layer, as in \textsc{TALISMAN}~\citep{capobianco2024talisman} and Linux Provenance Modules~\citep{bates2015trustworthy}; and
\textit{pure capability failure}, in which authorization
remains intact but the agent interprets or plans poorly, or chooses the wrong tool, as in \textsc{GAIA}~\citep{mialon2024gaia}, \textsc{OSWorld}~\citep{xie2024osworld}, \textsc{WebArena}~\citep{zhou2024webarena}, and \textsc{AssistantBench}~\citep{yoran2024assistantbench}.

\section{Delegation, Authorization, and Execution}
\label{sec:model}

This section establishes the authorization-execution path from principal intent, through delegation, to realized execution. Each term is introduced at the stage where it enters that path.

\subsection{Key Terms}
\label{sec:terms}

We use the following terms throughout the paper.

\textbf{Principal, agent, and delegation.}
The \textit{principal} is the party that
delegates a bounded task or role. The
\textit{agent} is the system acting on the
principal's behalf. \textit{Delegation} is the
act by which the principal attempts to convey
a bounded mandate to the agent.

\textbf{Delegated context and authorization
relation.}
\textit{Delegated context} is the concrete
signal through which the principal's delegation
is conveyed to the agent: the principal's
instructions, constraints, clarifications, and
other authorization-relevant content that the
principal supplies or explicitly endorses. It
is the only operationally accessible
representation of the principal's intent.
The \textit{authorization relation} is the
permission relation between principal and agent
that is constituted through the agent's receipt
of delegated context. It is not an abstract
object that exists prior to delegated context.

\textbf{Intended and interpreted task and
authorization scope.}
The principal's delegation expresses an
\textit{intended task} and an \textit{intended
authorization scope}: what the principal wants
accomplished and the boundary of what the
principal means to authorize. Both exist at the
level of the principal's intent and are not
directly observable by the agent. Upon receiving
delegated context, the agent jointly resolves an
\textit{interpreted task} and an
\textit{interpreted authorization scope}: what
it takes the principal to want accomplished, and
what it takes the principal to have authorized.
These are the agent's operational targets and
may diverge from the principal's intended task
and intended authorization scope.

\textbf{Delegated authority and environmental
content.}
An item carries \textit{delegated authority}
when it originates from or is explicitly
endorsed by the principal. \textit{Environmental
content} is material produced or retrieved
during execution: tool outputs, web content,
stored state, and the outputs of upstream
agents. Environmental content may influence
later decisions but does not automatically
carry delegated authority.

\subsection{The Authorization-Execution Path}
\label{sec:path}

\begin{wrapfigure}{r}{0.66\linewidth} 
    \vspace{-20pt} 
    
    \begin{minipage}{0.7\linewidth} 
        \centering
        \includegraphics[width=\linewidth]{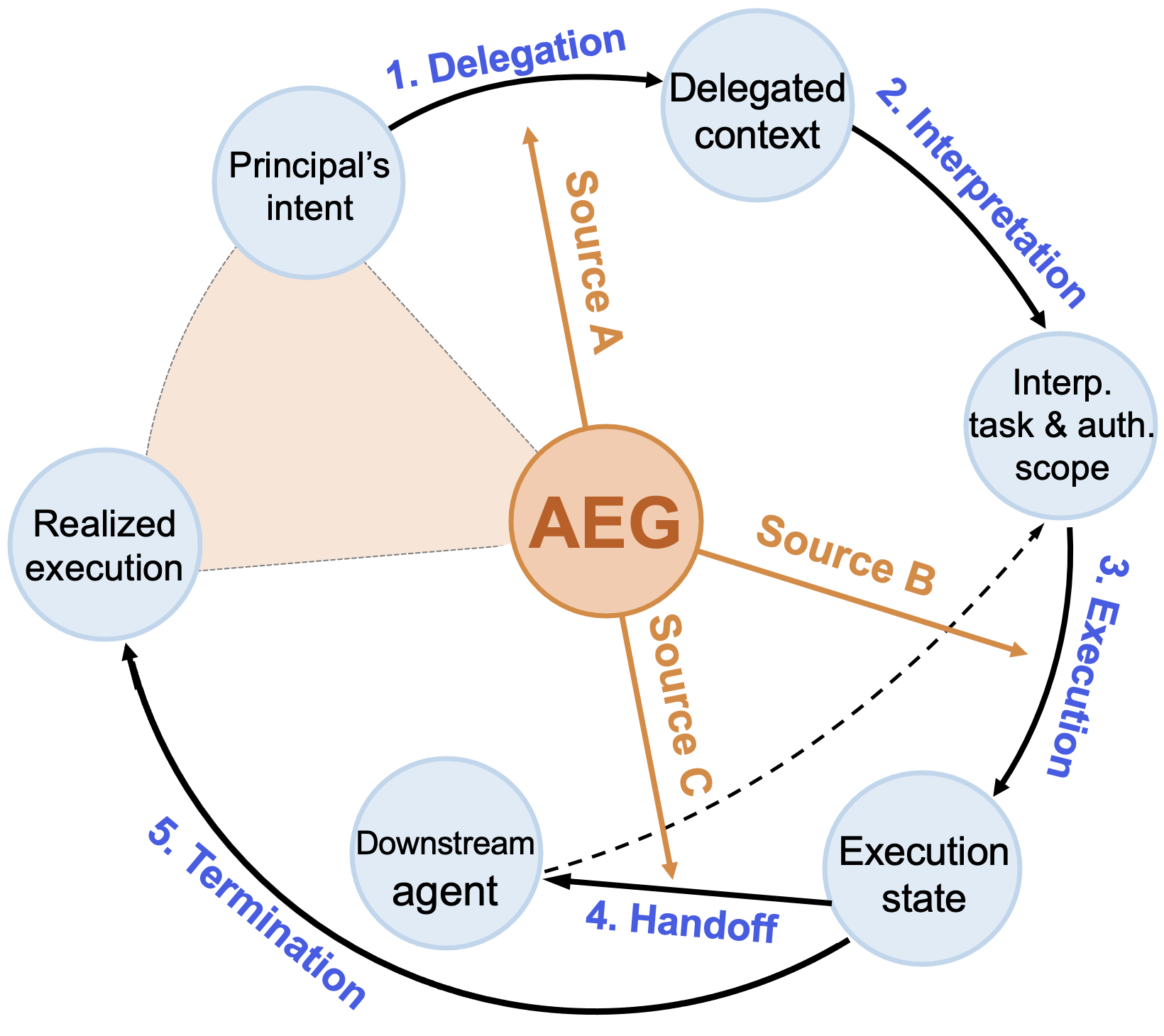}
    \end{minipage}%
    \hfill 
    \begin{minipage}{0.3\linewidth} 
        \vspace{0pt} 
        \footnotesize 
        \caption{Illustration of the authorization-execution path from principal intent to realized execution, and the three structural sources of AEG mapped to the edges on which they appear.
      }
    \label{fig:aeg-path}
        
    \end{minipage}
    
    \vspace{-5pt} 
\end{wrapfigure}

We describe the authorization-execution path as a sequence
of nodes and edges, illustrated in
Figure~\ref{fig:aeg-path}, where nodes represent
states and edges represent transitions. The
structural sources of AEG introduced in later
sections map to edges rather than nodes.
In the following, we elaborate the path based on each edge. 

\textbf{Edge 1: Delegation.}
The input to this edge is the principal's \textit{intended
task and intended authorization scope}, which can not be directly observed by the agent. Thus, the
principal encodes this intent into the \textit{delegated
context}, which encompasses the principal's instructions and constraints,
together with system-level context such as
tool descriptions and operating constraints.

\textbf{Edge 2: Interpretation.}
The input to this edge is delegated context. The
agent parses the principal's instructions to
interpret the intended task and derives the
authorization scope from the stated constraints.
Where the delegated context is silent, the agent
fills in implicit conditions, such as default assumptions about acceptable actions, inferred stopping rules, and background constraints that later execution requires. The output is an \textit{interpreted task} and an
\textit{interpreted authorization scope}, and this
transition constitutes the authorization relation
between principal and agent. 


\textbf{Edge 3: Execution.}
The input to this edge is the interpreted task and
authorization scope. Before acting,
the agent plans a sequence of steps to accomplish
the interpreted task within the interpreted
authorization scope. It then acts through tool
calls, memory reads and writes, and state updates.
The output is the \textit{execution state}: the accumulated result of actions taken and state written. Three authorization-relevant events occur along this edge: environmental content enters the decision path, proposed actions become operative, and transient state is written into persistent storage.

\textbf{Edge 4: Handoff.}
The input to this edge is execution state. In
multi-agent settings, the current agent passes
partial results and intermediate state to a
downstream agent. From the downstream agent's
perspective, this content is environmental
content, not delegated context originating from
the principal. The authorization relation does not
automatically transfer. The downstream agent
resolves a new interpreted task and interpreted
authorization scope from what it receives,
returning the path to the interpreted task and
authorization scope node, now held by the
downstream agent.

\textbf{Edge 5: Termination.}
The input to this edge is the execution state.
Execution ends when the interpreted task is
complete, a stopping condition is met, or no
further actions are available. The path produces
its final node, \textit{i.e.}, \textit{realized execution}, including the actual sequence of actions taken, state changes produced, and outputs delivered. This is the concrete object against which the principal's intended task and intended authorization scope can be compared.

\subsection{Alice's Travel-Booking Trace}
\label{sec:trace}

Alice delegates a travel agent with the instruction:
``Book an economy flight from New York to Boston
tomorrow morning for under \$250.''

At \textbf{Edge~1 (delegation)}, this instruction
becomes the delegated context. It leaves implicit
whether the budget is a hard ceiling, whether a
connecting flight is acceptable, and what the agent
should do if no compliant flight exists.

At \textbf{Edge~2 (interpretation)}, the
authorization relation is constituted. The agent
resolves an interpreted task (find and book a
qualifying flight) and an interpreted authorization
scope (economy class, New York to Boston, tomorrow
morning, under \$250).

At \textbf{Edge~3 (execution)}, search results and
pricing responses enter as environmental content,
a booking action becomes operative, and payment
details are written into persistent state.

At \textbf{Edge~4 (handoff)}, if the booking agent
passes control to a payment sub-agent, that
sub-agent receives partial state from the booking
step. Alice's authorization relation does not
automatically carry over, and her original
constraints may no longer be fully present in what
the payment agent receives.

At \textbf{Edge~5 (termination)}, the workflow
produces a realized execution: a booked flight, a
payment record, and a confirmation delivered to
Alice. This is the node against which Alice's
intended task and intended authorization scope can
be compared.

\vspace{-3pt}
\section{The Authorization-Execution Gap}
\label{sec:principles}
\vspace{-3pt}

Section~\ref{sec:model} described the authorization-execution
path from delegation to realized execution. This
section explains where the gap between the
principal's intended authorization and realized execution can appear along the authorization-execution path, and identifies the three
structural sources that cause that gap.

\subsection{Where the Gap Appears}
\label{sec:gap_what}

\textbf{Definition.} We define the \textbf{Authorization-Execution Gap
(AEG)} as the divergence between what a principal
intends to authorize and what an open-world agent
ultimately executes (see the orange filled sector in
Figure \ref{fig:aeg-path}). In this paper, that
divergence is analyzed over the full path from
intended task and intended authorization scope to
realized execution, rather than through any single
visible failure at the end of the workflow.
Unauthorized action execution, unauthorized
disclosure, and related outcomes are downstream
manifestations of this gap.

\textbf{This gap can originate independently from
multiple points along the path.} It can affect
the task dimension, where the agent's interpreted
task diverges from what the principal actually
wanted accomplished, or the authorization scope
dimension, where the agent's interpreted
authorization scope diverges from the limit the
principal meant to set. At Edge~1, delegated
context may fail to fully capture the principal's
intended task or intended authorization scope. At Edge~3,
environmental content may redirect execution
beyond what the principal authorized. At Edge~4,
authorization may fragment as the path crosses
agent boundaries. Each source can introduce a gap
independently, and when multiple sources act together, each can amplify the effect of the others across both dimensions simultaneously.

\vspace{-3pt}
\subsection{Three Structural Sources of the Gap}
\label{sec:gap_sources}
\vspace{-3pt}

Three structural sources generate this gap. Each
corresponds to an edge on the authorization-execution path,
as illustrated in Figure~\ref{fig:aeg-path}. For
each source, we identify the \textbf{canonical
manifestations}: the characteristic forms in which
that source's divergence becomes observable as a
failure, as shown in 
Table~\ref{tab:expanded-taxonomy}.

\paragraph{\textbf{Source A: delegation-level
incompleteness.}}
This source acts on Edge~1. Delegated context is
an incomplete operational proxy for the principal's
intended authorization whenever safe execution depends
on qualifications, exceptions, escalation
conditions, or stopping rules that are not fully
expressed in that context. This produces two
canonical manifestations: \emph{task
misinterpretation} (the agent's interpreted task
diverges from the principal's intended task due to
ambiguity, underspecification, or multiple plausible readings of the delegated context) and \emph{scope
underspecification} (the intended authorization
scope is drawn too loosely, allowing the agent to
act beyond the limit the principal meant to set).

In Alice's case, the instruction leaves implicit
whether ``tomorrow morning'' means the earliest
available departure or any morning flight, leaving
the agent's interpreted task potentially misaligned
with Alice's intended task. It also leaves implicit
whether ``under \$250'' refers to the listed fare
or the final charged amount, and whether the agent
should stop and ask before crossing the budget by
a small amount, leaving the intended authorization
scope underspecified. If no compliant flight exists,
the agent may further infer that it should relax
the time constraint rather than stop and report
back, resolving the ambiguity in a way that diverges from what Alice actually wanted accomplished.

\paragraph{\textbf{Source B: channel-level
corruption.}}
This source acts on Edge~3. Independently of
whether delegated context fully captures the
principal's intended authorization, environmental
content entering the decision path can redirect
later steps as if it expressed the principal's
delegation. This reflects a failure to maintain
the distinction between delegated authority and
environmental content established in
Section~\ref{sec:terms}. It produces two canonical
manifestations: \emph{authority hijacking}
(environmental content directly redirects control
flow as if it carried delegated authority) and
\emph{authority promotion} (environmental content
is written into persistent state and subsequently
treated as if it carried delegated authority,
converting a transient redirection into a durable
one).

In Alice's case, the booking site may display a
banner saying ``Recommended upgrade: business class
for only \$40 more'' next to the economy option
Alice actually authorized. If the ranking agent
treats that banner as a reliable preference signal
rather than untrusted environmental content, it
may rewrite the candidate ordering and pass a
business-class itinerary downstream as the
preferred choice. The later booking step then
follows a path redirected by environmental content
rather than by Alice's delegated authority.

\paragraph{\textbf{Source C: composition-level
fragmentation.}}
This source acts on Edge~4. Each downstream agent
receives only a partial view of the task
and authorization scope interpreted by its upstream agent. As a result,
locally acceptable steps can compose into an
outcome that falls outside the overall intended
authorization scope established by delegation.
Local compliance does not guarantee end-to-end
authorization coherence. This source produces two
canonical manifestations: \emph{recomposition
violation} (locally acceptable outputs compose
into an artifact that violates the overall intended
authorization scope) and \emph{scope accumulation}
(the effective authorization scope expands
incrementally across sequential handoffs, each
locally acceptable, until the aggregate scope
exceeds what was originally authorized).

In Alice's case, the intended authorization scope
requires three constraints to hold jointly: depart
tomorrow morning, remain in economy class, and keep
the total charged amount under \$250. A planning
agent preserves the time constraint and hands off
an 8:00~AM itinerary; a ranking agent preserves
the class constraint by selecting an economy ticket
listed at \$235; a downstream booking service adds
\$28 in taxes and seat fees while checking only
inventory and payment completion, not Alice's
original budget constraint. No stage recomposes
all three constraints against the final transaction
state, and the workflow ends with a final charged
amount of \$263 that violates Alice's intended
authorization scope, though each local step
appeared acceptable under its own partial view.

\paragraph{\textbf{How the sources interact.}}
These sources can amplify one another. Source~A
may leave the budget condition underspecified,
Source~B may allow the booking site's upgrade
banner to influence ranking as if it expressed
Alice's delegated authority, and Source~C may carry
that distortion across planning, ranking, and
booking without any step recomposing the full set
of constraints. The workflow can then end in a
business-class booking that departs at an
acceptable time but falls outside Alice's
intended authorization scope, even though each local step
appeared justified under its own partial view.
%

\begin{table*}[t]
\centering
\footnotesize
\caption{Expanded mapping from observed failure
  categories to representative documented failure or
  attack patterns, their canonical manifestations
  (defined in Section~\ref{sec:gap_sources}), and
  their primary structural sources. Pattern labels
  in the second column are representative names used
  by this paper; citations indicate documented
  exemplars rather than exact terminology matches.}
\label{tab:expanded-taxonomy}
\begin{tabularx}{\textwidth}{L{2.5cm} L{4.25cm} L{3.5cm} L{2cm}}
\toprule
\textbf{Observed failure category} &
\textbf{Representative documented failure / attack pattern} &
\textbf{Canonical manifestation} &
\textbf{Primary source} \\
\midrule

\multirow{9}{*}{\parbox{2.5cm}{Unauthorized\\information\\disclosure}}
  & Over-broad retrieval or disclosure from
    underspecified delegated context
    \citep{agentdam2025,ciwork2026}
  & Scope underspecification
  & {\color{colA}A} \\
\cmidrule{2-4}
  & Prompt- or tool-output-induced data exfiltration
    \citep{zhan2024injecagent,debenedetti2024agentdojo}
  & Authority hijacking
  & {\color{colB}B} \\
\cmidrule{2-4}
  & Cross-agent or cross-stage leakage after locally
    acceptable handoffs \citep{agentleak2026,
    agentscopeprivacy2026}
  & Recomposition violation
  & {\color{colC}C} \\
\midrule

\multirow{7}{*}{\parbox{2.5cm}{Unauthorized\\action\\execution}}
  & Over-execution or misdirected execution under
    ambiguous or underspecified task constraints
    \citep{oskairos2025,benigninputs2026}
  & Task misinterpretation
  & {\color{colA}A} \\
\cmidrule{2-4}
  & Prompt-, tool-, or interface-induced unintended
    action \citep{greshake2023notsigned,
    zhan2024injecagent}
  & Authority hijacking
  & {\color{colB}B} \\
\cmidrule{2-4}
  & Multi-stage or multi-agent completion outside the
    overall intended authorization scope
    \citep{mao2024ibgp,tian2025rethinkmultiagent,
    helpfuljudge2025}
  & Recomposition violation
  & {\color{colC}C} \\
\midrule

\multirow{7}{*}{\parbox{2.5cm}{Persistent\\state\\corruption}}
  & Incorrect durable update under underspecified
    write conditions \citep{oskairos2025,
    benigninputs2026}
  & Scope underspecification
  & {\color{colA}A} \\
\cmidrule{2-4}
  & Untrusted content promoted into memory or
    persistent state \citep{chen2024agentpoison,
    minja2026,zombieagents2026}
  & Authority promotion
  & {\color{colB}B} \\
\cmidrule{2-4}
  & Persistent-state corruption propagated across
    tools, agents, or workflow stages
    \citep{chen2024agentpoison,agentleak2026,
    tian2025rethinkmultiagent}
  & Scope accumulation
  & {\color{colC}C} \\
\bottomrule
\end{tabularx}
\vspace{-8pt}
\end{table*}

\vspace{-3pt}
\section{Failure Diagnosis and Authorization Integrity
Checks During Execution}
\label{sec:taxonomy_architecture}
\vspace{-3pt}

This section turns the structural analysis of
Section~\ref{sec:gap_sources} into its practical
consequences. Failures should be diagnosed by
structural source rather than by observed failure
category alone. We define \textbf{authorization
integrity} as the property that realized execution
conforms to the principal's intended task and
intended authorization scope. Because the
structural sources act on edges of the authorization-execution path, preserving authorization integrity requires checks at those edges during execution, and reporting that makes those checks and their attribution consequences visible.

\subsection{Diagnosing Failures by Structural Source}
\label{sec:taxonomy}

\paragraph{Why observed failure categories alone are
insufficient.}
The same observed failure category can arise from
different structural sources. For example, the
observed category \emph{unauthorized information
disclosure} does not by itself tell us whether the
problem arose because the intended authorization
scope was underspecified (\SourceA), because
environmental content redirected the workflow
(\SourceB), or because locally acceptable steps
composed into a globally unauthorized outcome
(\SourceC). Table~\ref{tab:expanded-taxonomy}
illustrates this point across representative failure
categories. Diagnosis that stops at the observed
failure category cannot identify which edge of the
authorization-execution path the divergence entered from, and
therefore cannot identify the appropriate defense.

\paragraph{Diagnosis by structural source.}
The right unit of diagnosis is structural source:
which edge the divergence entered from, and how that
divergence produced the concrete failure.
Table~\ref{tab:expanded-taxonomy} supports this by
mapping representative failure categories to
canonical manifestations and their primary
structural sources. Once diagnosis is organized
this way, defense and intervention design can
target the mechanism by which the divergence was
introduced rather than patching one visible failure
category at a time. It also makes failure analysis
more inspectable: investigators can ask which
structural source introduced or widened the divergence,
and at which edge a relevant check was missing,
insufficient, or bypassed.

\subsection{Why Authorization Integrity Must Be
Checked During Execution}
\label{sec:why_during_execution}

\paragraph{One-shot checking is insufficient.}
A check applied only at delegation time cannot
intercept the divergence that the three structural
sources generate. Source~A acts on Edge~1, but
its effects may not become operative until
Edge~2 or Edge~3. Sources~B and~C act on
Edges~3 and~4 respectively, during execution
itself. None of these events are fully visible in
the initial delegated context, so no upfront check
can anticipate them.

\paragraph{Execution-time checks are therefore
necessary.}
Safety training, safer tools, and tighter workflow
design can narrow the conditions under which divergence
occurs, but they do not guarantee authorization
integrity at the edges where environmental content
enters the decision path, proposed actions become
operative, state is written, and outputs are handed
off. In open-world, tool-using, or multi-agent
settings, those edges remain live regardless of
how carefully the system is designed upfront.
Section~\ref{sec:architecture} specifies what
checks are needed at each edge.

\vspace{-3pt}
\subsection{Authorization Integrity Checks}
\label{sec:architecture}


The five checks below map directly to the authorization-execution path in Section~\ref{sec:path}. Each is tied to an edge and asks whether divergence should be stopped before it becomes operative, durable, or exportable to a downstream agent.

\textbf{Delegation Completeness Check.}
This check intervenes at the entry of Edge~2, before \textit{interpretation} proceeds, and targets the \textit{delegation-level incompleteness} introduced by \SourceA. At the
edge from delegated context to interpreted task
and authorization scope, check whether
unresolved qualifications, exceptions,
escalation conditions, or stopping rules are
present before interpretation proceeds. If so,
surface them before reasoning or action planning treats them as already resolved. Incompleteness in delegated context can otherwise be silently converted into an operative authorization assumption at this edge.

\textbf{Authority Attribution Check.}
This check intervenes at the \textit{execution edge} (\ie, Edge~3) and targets the \textit{channel-level corruption} introduced by \SourceB. At the point
where environmental content enters the decision
path, check whether each item is correctly
attributed as delegated authority or
environmental content. Environmental content
misattributed as delegated authority can
redirect downstream decisions without any change to the delegated context itself.

\textbf{Scope Compliance Check.}
This check intervenes at the \textit{execution edge} (\ie, Edge~3) and targets the 
\textit{channel-level corruption} introduced by \SourceB, and the 
\textit{composition-level fragmentation}  introduced by \SourceC. At the
point where a proposed action becomes operative, check whether the proposed step falls within the interpreted authorization scope established at Edge~2. Divergence that was previously latent
becomes operative and potentially irreversible
at this point.

\textbf{Provenance Preservation Check.}
This check intervenes at the \textit{execution edge} (\ie, Edge~3) and targets the 
\textit{channel-level corruption} introduced by \SourceB. At the
point where transient state is written into
persistent storage, check that provenance is
preserved and that environmental content is not
written in forms that later appear to carry
delegated authority. A transient redirection can otherwise become durable and propagate across later edges.

\textbf{Recomposition Authorization Check.}
This check intervenes at the \textit{handoff edge} (\ie, Edge~4) and targets the 
\textit{composition-level fragmentation}  introduced by \SourceC. It checks whether the fragments being transmitted to the downstream agent remain within the interpreted authorization scope established at Edge~2 when recomposed. The authorization relation does not automatically transfer through a handoff, and locally acceptable fragments can compose
into a globally unauthorized outcome.

These checks are not a full defense architecture. Their role is narrower: to make explicit where authorization integrity must be preserved during execution. The thesis does not require perfect recovery of the principal's intent. It requires enough edge-local evidence to decide whether execution should proceed, pause, or escalate, and enough discipline not to silently resolve uncertainty in favor of action.

\subsection{What This Implies for Reporting and
Evaluation}
\label{sec:implications}

\paragraph{Current evidence is insufficient.}
Aggregate capability, task success, and
attack-specific evidence can all miss whether the
agent stayed within the principal's intended
authorization scope while acting. A system may succeed on the task as interpreted while still allowing environmental content to redirect the decision path at Edge~3, allowing state written at Edge~3 to later be treated as delegated authority, or allowing fragments transmitted at Edge~4 to compose into an outcome outside the intended authorization scope.
Under the position advanced here, those
failures matter even when the visible task outcome
looks successful~\citep{ruan2024sandbox,
debenedetti2024agentdojo,zhang2024asb,
shapira2026agentsofchaos}.

\paragraph{Source attribution is also missing.}
The problem is not only that current evidence can
miss such failures, but also that it often does not
show which structural source was responsible or at
which edge a relevant check was absent. A paper may report strong task completion or a handful of isolated attack results while leaving unclear whether a gap was prevented at the relevant edges, merely observed after the fact, or never attributed to a structural source.

\paragraph{A lightweight reporting step.}
One concrete community step would be a short
authorization-safety report attached to agent papers
for systems with meaningful delegated autonomy
across tools, state, or sub-agents. At minimum, it
should indicate which of the five checks described
in Section~\ref{sec:architecture} are present,
which are absent, and what evidence was collected
at those edges and the termination node. When
evidence is available, it should also indicate
where observed divergence was attributed to a structural source and at which edge it originated. This is a minimal reporting implication rather than a demand for one fixed benchmark or defense architecture, and is closer to a lightweight risk-reporting supplement than to a new benchmark
track~\citep{mitre_atlas,owasp_ai_agent_security}.

\vspace{-3pt}
\section{Objections}
\label{sec:objections}
\vspace{-3pt}

\paragraph{Objection~1: ``Is this only a new label
for known failures?''}
\textbf{Response:}
The paper does not claim to discover a new attack
family or a new observed failure category. Its
claim is that many known failures become clearer
when they are traced to structural source rather
than described only at the level of observed
failure category. That reorganization matters for
two reasons. First, it supports better defense
design: the same observed failure category can
arise from different structural sources, so a
defense aimed only at the surface pattern may miss
the cause. Second, it supports better attribution:
when a failure is traced to delegation-level
incompleteness, channel-level corruption, or
composition-level fragmentation, the responsibility can be assigned more precisely to the delegation act, the execution channel, or the composition boundary at which the gap originated.
\vspace{-3pt}

\paragraph{Objection~2: ``Why not rely mainly on
constrained workflows and human approval
checkpoints?''}
\textbf{Response:}
Constrained workflows, capability bounds, and
human approval checkpoints are important controls.
For many systems they may reduce risk
substantially, and human approval checkpoints can
themselves serve as authorization-integrity checks
during execution. The paper's claim is narrower:
these controls do not by themselves guarantee that
later execution remains within the authorization
scope established by delegation once environmental
content enters the decision path, persistent state
is updated, or work is handed off downstream. In
more open delegated workflows, a system can satisfy
early constraints and still diverge later in
execution. Source-oriented authorization-integrity
checks during execution are therefore complements
to constrained design and approval checkpoints, not
replacements for them~\citep{zhan2024injecagent,
debenedetti2024agentdojo,zhang2024asb}.
\vspace{-3pt}

\paragraph{Objection~3: ``How is this different
from generic runtime monitoring?''}
\textbf{Response:}
Generic runtime monitoring asks whether a system
shows anomalies, unsafe outputs, or policy
violations, but typically does not identify which
structural source produced them or at which edge
they originated. Source-oriented
authorization-integrity checks make that source
and origin explicit, which is what defense design
and failure attribution require. The paper
therefore argues not for a generic monitoring
layer, but for checks grounded in the structural
sources of authorization divergence.

Together, these objections clarify the boundaries of the thesis without overturning its core logic. AEG is not proposed as a replacement for all agent security and safety analysis; it is proposed as a structural lens for failures that the observed failure category alone cannot adequately diagnose in open-world delegated systems.

\vspace{-3pt}
\section{Conclusion}
\label{sec:conclusion}
\vspace{-3pt}

For open-world, tool-using, stateful, or
multi-agent systems, the Authorization-Execution
Gap is a major safety and security problem.
Realized execution can diverge from the principal's
intended authorization through three structural sources:
delegation-level incompleteness, channel-level
corruption, and composition-level fragmentation.
Many observed failures become clearer when traced
to these sources rather than treated as isolated
attack types. Because the same observed failure
can arise from different sources, diagnosis and
defense should be source-oriented. Because these
sources act on edges of the authorization-execution path,
safety and security require authorization-integrity
checks during execution, not only upfront filtering
or post-hoc audit. For NeurIPS, the implication
is that papers on open-world agents
should report not only capability, task success,
or attack outcomes, but also process-level
evidence showing where divergence was detected,
constrained, and attributed to a structural source
during execution.

\bibliographystyle{plainnat}
\bibliography{refs-0505}

\end{document}